\documentclass[journal]{IEEEtran}
\usepackage[numbers]{natbib}
\usepackage{amsmath}
\usepackage[gen]{eurosym}

\usepackage[labelformat=simple]{subcaption}

\DeclareCaptionLabelFormat{subcaptionlabel}{\normalfont(\textbf{#2}\normalfont)}
\ifCLASSINFOpdf
  \usepackage[pdftex]{graphicx}
\else
\fi
\begin{document}
%
\title{Optimal Scheduling of Flexible Power-to-X Technologies in the Day-ahead Electricity Market}
%
%
%

\author{Neeraj~Dhanraj~Bokde,
        Tim~T~Pedersen, and~Gorm~Bruun~Andresen
\thanks{N. D. Bokde, T. T. Pedersen, and G. B. Andresen were with the Department of Mechanical and Production Engineering - Renewable Energy and Thermodynamics, Aarhus University, 8000, Denmark and iCLIMATE Interdisciplinary Centre for Climate Change, Aarhus University, Aarhus, Denmark. e-mail: gba@mpe.au.dk}
}

\maketitle

\begin{abstract}
The ambitious CO$_2$ emission targets of the Paris agreements are achievable only with renewable energy, CO$_2$-free power generation, new policies, and planning. The main motivation of this paper is that future green fuels from power-to-X assets should be produced from power with the lowest possible emissions while still keeping the cost of electricity low. 
To this end we propose a power-to-X scheduling framework that is capable of co-optimizing CO$_2$ emission intensity and electricity prices in the day-ahead electricity market scheduling. Three realistic models for local production units are developed for flexible dispatch and the impact on electricity market scheduling is examined. Furthermore, the possible benefits of using CO$_2$ emission intensity and electricity prices trade-off in scheduling are discussed. We find that there is a non-linear trade-off between CO$_2$ emission intensity and cost, favoring a weighted optimization between the two objectives. 
\end{abstract}


%
\IEEEpeerreviewmaketitle

\section{Introduction}
%
%
%
%
\IEEEPARstart{I}{n}  recent years, the world has paid enormous attention to the effects caused by greenhouse gas emissions and climate changes \cite{goglio2020advances}. It is estimated that the global temperature will rise on average 1.3~$^\circ$C as compared to that of the pre-industrial revolution period and greenhouse gases are the major factors responsible for it \cite{seixas2020carbon}.

Most countries have already shown their commitments to reduce CO$_2$ emissions at several international platforms such as the United Nations Climate Change Conference at Copenhagen in 2009 and the Paris Agreement in 2015. These commitments were towards reducing CO$_2$ emissions, increasing renewable energy shares and efficiencies for the whole world. Such commitments are achievable only with practical CO$_2$ free power generations and solar, wind, and other renewable power sources based energy systems \cite{bekun2019toward}. Therefore, new policies and planning are mandatory to manage climate change by reducing renewable energy costs and the decarbonization of the energy systems \cite{mangla2020step, razmjoo2020technical}.

To fulfill such ambitious targets, it proves urgent for the world, as an international community, to search for effective and efficient options to control as well as reduce carbon emissions, and to maintain a good balance between economic growth and carbon emissions. 

The majority of electricity produced and consumed in Europe is traded in the European power markets. Power trading is done in several markets on different time scales. The majority of electricity is traded in the Day-ahead market. The Day-ahead market clears at noon the day before delivery and determines the hourly prices of electricity for the next day. These prices are based on matching bids and offer received from producers and consumers. When bidding in the electricity market, one needs to specify the amount of energy in MWh for a specific hour slot to be purchased or sold at a certain price level (marginal cost in \euro{}/MWh) for the following day.

\begin{figure}[h]
	\centering{\includegraphics[width=0.49\textwidth]{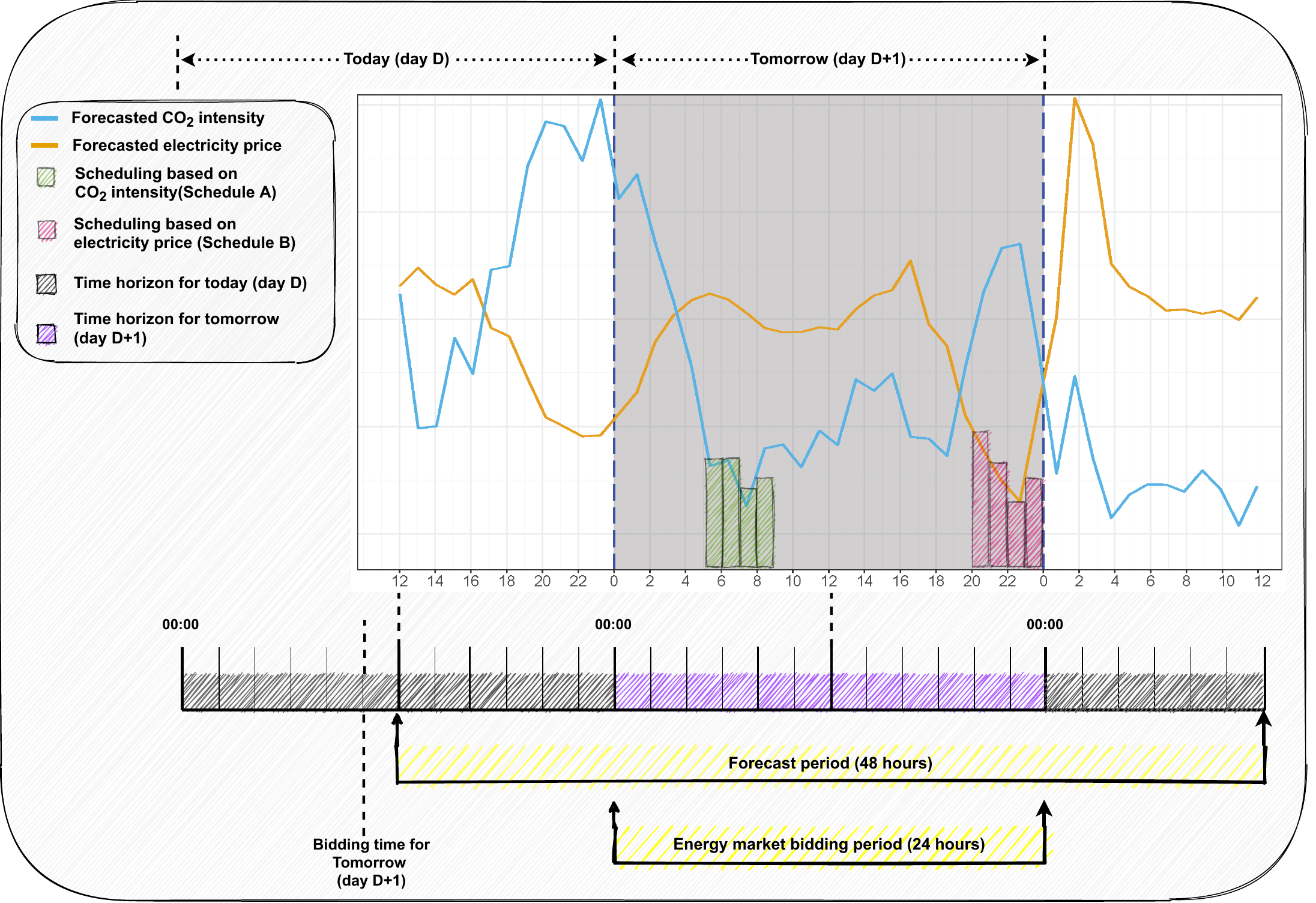}}
	\caption{Time framework to forecast day-ahead energy market and example of scheduling 4 hours of electricity consumption in western Denmark (DK-1) based on electricity prices and CO$_2$ intensity.}
	\label{Fig_1}
\end{figure}

The operations of Day-ahead markets and bidding are demonstrated with a time framework shown in Figure \ref{Fig_1}. In this figure, the day `$D$' represents today (shown in a black strip), on which the consumer needs to bid for the following day (`$D+1$') before noon on the day `$D$'. Ideally, the bidding period lies between $09.00 - 10.00$ AM. The decision of bidding is generally based on the accuracy of electricity prices forecasts in the day-ahead energy market. These forecasted values for tomorrow (`$D+1$') is of interest to the market auctions.

Currently, the predominant factor for bidding is the power price. However, recent work has presented methods capable of predicting the CO$_2$ intensity of the power \cite{bokde281short}. In this work, we will utilize this to formulate a scheduling algorithm that considers emissions along with the price. 
Electricity prices and CO$_2$ intensity values are shown in Figure \ref{Fig_1} are the forecasted values for 48 hours with the ‘Method 1’ proposed in \cite{bokde281short}. 
A minimum of 36 hours should be forecasted, as bids are placed at 12:00 for the entire period of 00:00-24:00 the next day. Thus the forecasted values from 12:00-24:00 on day `$D$' are not used. In Figure \ref{Fig_1}, the forecasted values for the day `$D+1$' (i.e., tomorrow, shown in a violet stripe), are the only values from the forecast used for scheduling. 
Based on the forecasted values in Figure \ref{Fig_1}, four hours are scheduled per minimum values of CO$_2$ intensity (Schedule A, shown in green) and electricity prices (Schedule B, shown in violet). The hours selected in Schedule A indicate the electricity per hour responsible for minimum carbon emissions on the day `$D+1$'. Similarly, Schedule B indicates four hours having cheaper electricity costs on day `$D+1$'.

It is worth noticing that the hours scheduled with schedules A and B are different for the targeted day. 
Acknowledging that low CO$_2$ intensity also has a value, scheduling of power consumption becomes a multi-objective optimization problem. The two objectives are to minimize cost and to minimize CO$_2$ intensity.

Our earlier study \cite{bokde2020graphical} proposed a graphical method to obtain a trade-off between Schedule A and B for energy markets in several European countries. 
In the study, we found varying correlations between CO$_2$ intensity and power price depending heavily on the country investigated. In most countries, there was a significant benefit to be obtained by co-optimizing CO$_2$ intensity and power price.
The countries, where patterns of electricity prices and CO$_2$ intensity exhibit weaker correlations are more likely to benefit from investigating the trade-off between schedules A and B. On contrary, the benefit is less significant in countries having strong correlations in patterns of electricity prices and CO$_2$ intensity. For example, sample schedules A and B for France is shown in Figure 4 in \cite{bokde2020graphical}. The hours selected with schedules A and B are almost identical and therefore provide little room to obtain an extra benefit from co-optimization. The detailed trade-off analysis for several European energy markets is discussed in Figures 9, 10, 11, and Table A.4 in \cite{bokde2020graphical}. Although the graphical approach helped estimate the day-ahead market hours, which are responsible for cheaper electricity rates with fewer carbon emissions, its utility was limited. With this approach, it was not possible to consider the user-specific electricity generating technologies in the spot market systems. Besides, the methodology was dedicated to very specific targets without any external constraints. To overcome these limitations and to provide extra flexibility to the existing frameworks of the spot market, a power-to-X scheduling framework is proposed in this paper. The detailed schematic of the proposed methodology is shown in Figure \ref{Fig_3}.

In the present study, we considered the Danish energy market for carbon and cost-efficient scheduling for power-to-X applications. Application to several Power-to-X technologies are is tested. 

The objectives of the presented study are:
\begin{itemize}
	\item To develop a power-to-x scheduling framework capable of co-optimizing CO$_2$ intensity and price in the scheduled hours while considering technical limits of the plant operation. 
	\item Develop a testing framework, using the Danish power markets allowing us the quantify the obtain benefits.
	\item To study the possible benefits of using a CO$_2$ intensity/price trade-off in scheduling.
\end{itemize}

The rest of the paper is organized as follows. Section 2 describes the proposed methodology and technologies used in the study. Besides, this section describes the scheduling strategy proposed in this paper. The results and the performance evaluation are discussed in Section 3. The conclusions are presented in Section 4.

\section{Methodology}


As the share of VRE increase, periods with high availability of green electricity become more frequent. Thus, allocating a manageable demand in these hours can significantly lower the CO$_2$ emissions realized from the consumed power. In a similar way, demand located in times of low VRE generation may help to sustain the economy of fossil fuel generators such as coal power plants that produce low cost power during these periods.

In the present study, a power-to-X scheduling framework is proposed that is capable of co-optimizing CO$_2$ emissions and electricity prices. The scheduler operates in the day-ahead market, using the Danish electricity price zone DK-1 as reference. 
The proposed method can schedule any flexible demand plant, such as Power-to-X units, heat pumps, or more complex systems composed of several technologies. 

\begin{figure*}[h]
	\centering{\includegraphics[width=0.7\textwidth]{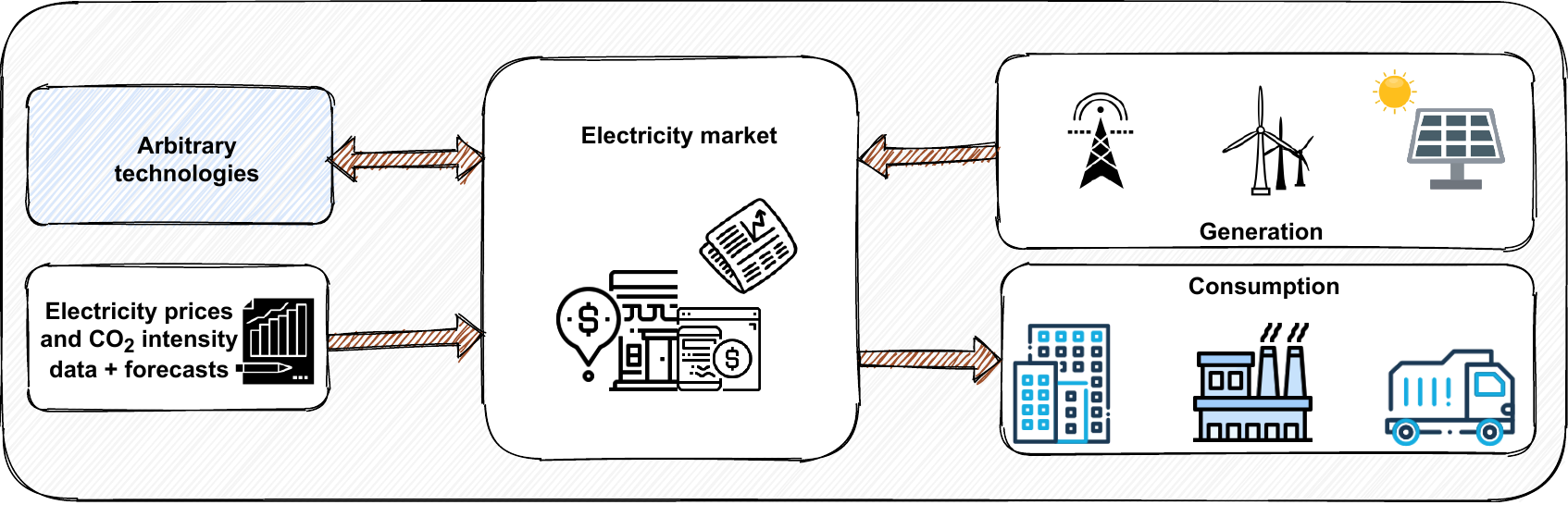}}
	\caption{Framework of integration of electricity market with AC grid, electricity loads and arbitrary technologies those are considered in the proposed study.}
	\label{Fig_2a}
\end{figure*}

The scheduling method consists of three separate modules; a set of forecasting algorithms, a scheduler, and a controller. 
The forecasting algorithms provide a 36-hour forecast for the spot price and CO$_2$ intensity of the electricity every day just before 12:00. Method 1 from the paper \cite{bokde281short} is used to forecast spot price and the method presented in \cite{tranberg2019real} is used to calculate CO$_2$ intensity of the electricity. Based on the forecasted values, the scheduler will determine the optimal hours to bid in the Day-ahead market. The scheduler is implemented in the open-source tool PyPSA \cite{PyPSA}. By modeling the technology to be scheduled in PyPSA, technical limitations such as ramping limits, minimum generation, etc. can be considered. The goal of the scheduler is to co-optimize the price paid for electricity and  CO$_2$ intensity from the purchased electricity, given the desired number of full load hours. The controller will analyze the realized bids and determine how many full load hours the plant should aim for the next day. The process of forecasting, scheduling, and updating the controller is repeated every day. A schematic of the setup can be seen in Figure \ref{Fig_2a}). Throughout this work, the technology is modeled as a price-taker, thus not affecting the spot price. 



\subsection{Power-to-X Technologies}
Power-to-X is considered to be one of the cornerstones in the transformation to a fully renewable-based energy system \cite{victoria2020early}. Generally, a power-to-X system converts electricity to another energy carrier, such as heat, hydrogen, methane, etc. This allows non-electric energy needs to be covered by green electricity. Furthermore, if the conversion process can be made flexible then power-to-X can also serve as a balancing unit in the power grid.



\begin{figure}[h]
	\centering{\includegraphics[width=0.44\textwidth]{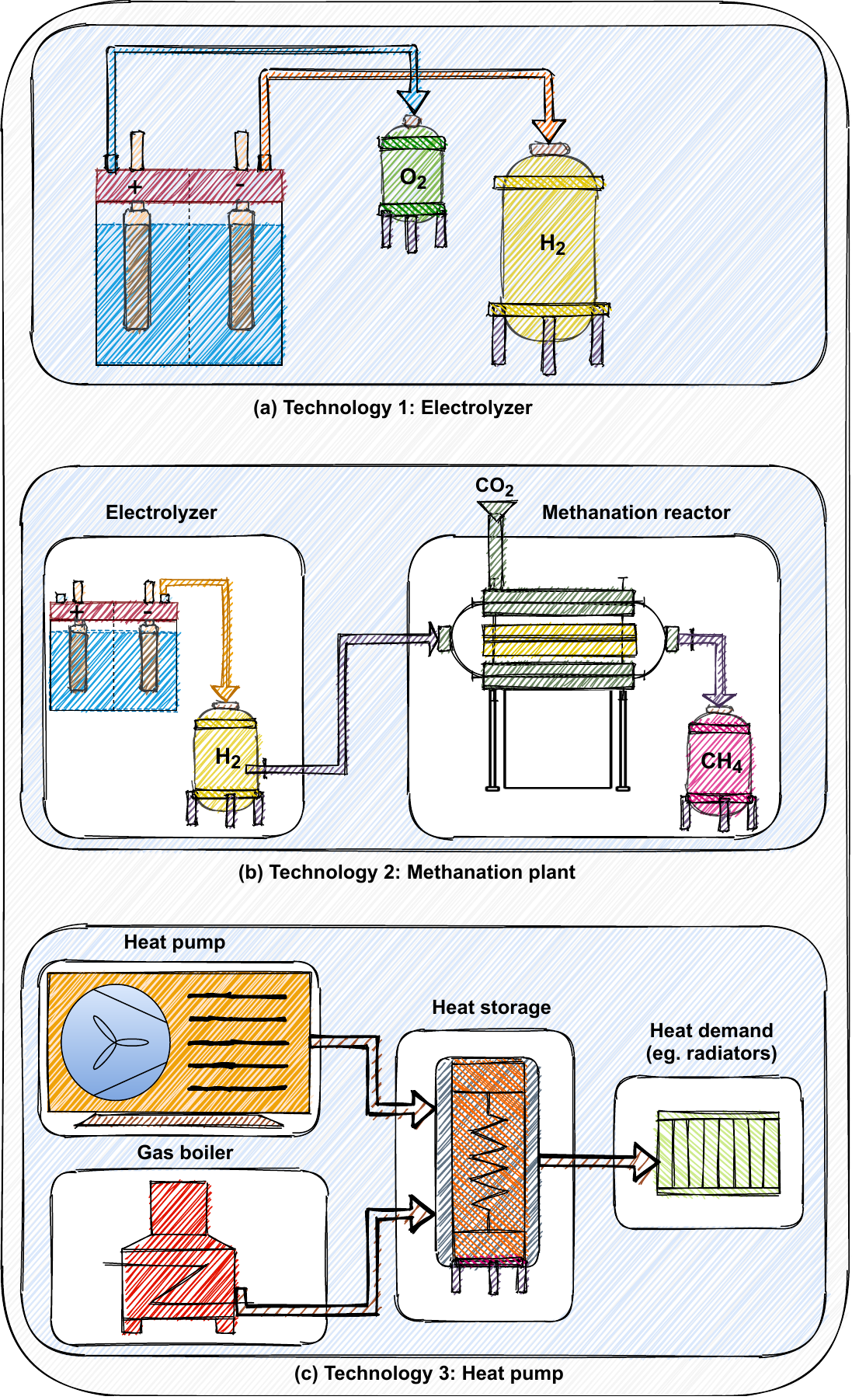}}
	\caption{Schematics of the power-to-x technologies modeled in this study.}
	\label{Fig_2b}
\end{figure}

In this study, three different power-to-X technologies of varying complexity are modeled. The first technology is a stand-alone electrolyzer. 
The second plant modeled is a Methanation plant. The plant consists of an electrolyzer cascaded with a methanation reactor. 
Last a heat pump combined with a gas boiler and thermal storage tank is modeled. A schematic of the technologies can be seen in Figure \ref{Fig_2b}.



\subsubsection{Electrolyzer}
Electrolyzers come in many shapes and sizes with different characteristics. The most prominent technologies currently are PEM, Alkaline, and SOC electrolyzers \cite{technologyCatalog}. On a high level, the working principle of the technologies is the same: Water is split into hydrogen and oxygen using electricity. The different types of electrolyzers have different constraints on ramping time, minimum load, idle power consumption, etc. Technical details of the electrolyzers are beyond the scope of this paper. 

The electrolyzer's role in the electricity system is as a consumer, with the capability of regulating consumption relatively fast. Thereby, the electrolyzers are capable of providing a service in the power markets as they can act as regulating power by varying production. Ideally, the electricity used for the electrolyzer operations should be delivered from renewable energy sources such as wind, solar, or hydro energy. Furthermore, as demand for green hydrogen increases, the CO$_2$ intensity of the hydrogen produced must be documented, thus, requiring the electrolyzer to operate in hours of low CO$_2$ intensity in the power grid. 

Electrolyzers play an important role in many power-to-X systems as many of these plants require a source of green hydrogen. This has also lead to a rapid development of the technology. A different application of electrolyzers is in a hydrogen storage facility. Here hydrogen is produced in the electrolyzer, stored in tanks, and then converted back to electricity in a fuel-cell when needed \cite{chen2009progress}. Using hydrogen as an energy carrier still has come challenges to overcome, and demand for green hydrogen will most likely initially be from industry applications.



Considering the benefit potentials and importance of electrolyzers in power-to-X applications, the first two technologies used in the present study are based on them. 
The first technology is a stand-alone electrolyzer that purchases electricity from the day-ahead market and generates hydrogen sold at a fixed price. A schematic is shown in Figure \ref{Fig_2b}(a)). The electrolyzer model with the efficiency of 70\% is employed \cite{budischak2013cost}, which is having a part load in sync with its operations in the last hours of the earlier day. Also, different ramp rates are assigned to the electrolyzer model.
The model parameters used for the electrolyzer are tabulated in Table \ref{T1}.

\subsubsection{Methanation plant}
The second technology used in this study is a cascade of an electrolyzer and a methanation plant (shown in Figure \ref{Fig_2b}(b)). 
The methanation plant modeling is based on the Sabatier process \cite{lunde1974modeling}.
In this model of a methanation plant, the methanation reactor consumes hydrogen generated and then stored by the electrolyzer operations. Besides, the required CO$_2$ for the methanation reactor is extracted from the flue gas, from the atmosphere, or through other sources. The model parameters (inspired from \cite{skov2019biogas}) used for the Methanation plant are tabulated in Table \ref{T2}.

\subsubsection{Heat pump, boiler, and storage tank}
The third technology model employed in the study is a heat pump coupled with heat storage along with a gas boiler. The schematic of the employed setup is shown in Figure \ref{Fig_2b}(c).
In this model, the electricity from the grid is converted into heat using the heat pump. The heat pump is assumed to have a Coefficient of Performance (COP) of 3. Along with a heat pump, the employed model consists of a gas boiler with 90\% efficiency, fixed gas prices and fixed CO$_2$ emissions, and heat storage of 90\% efficiency.
The model parameters used for heat pump and gas boilers are tabulated in Tables \ref{T3} and \ref{T4}, respectively. The values of these parameters are inspired from \cite{danish}.

This model is operated with some conditional procedure. A uniform heat load is applied to the unit. Furthermore, by default electricity is converted into heat through a heat pump and stored in the heat storage, which can directly be delivered to consumers. The heat pump operation is backed up with a gas boiler, in such a way that, when there is no heat available in the heat storage or when the marginal cost of the heat pumps is higher than a threshold, the gas boiler is used to produce heat and fill the heat storage. This marginal cost is based on the cost function shown in \eqref{e1a}. For the operations of the gas boiler, fixed gas prices and CO$_2$ emissions are allotted.

\subsection{Forecasting and CO$_2$ intensity estimation}
The aim of this paper is to perform optimal scheduling of flexible power-to-X technologies with the co-optimization of CO$_2$ emission intensity and electricity price values. This method is a data-driven approach and is based on 36-48 hours ahead forecasted values of hourly CO$_2$ emission intensity and electricity prices time series. I.e., two forecasts are needed. One forecast for CO$_2$ emission intensity in the grid, and one for the electricity price. 
This section will further discuss the nature of the forecast approach and models.

\begin{figure}[h]
	\centering{\includegraphics[width=0.5\textwidth]{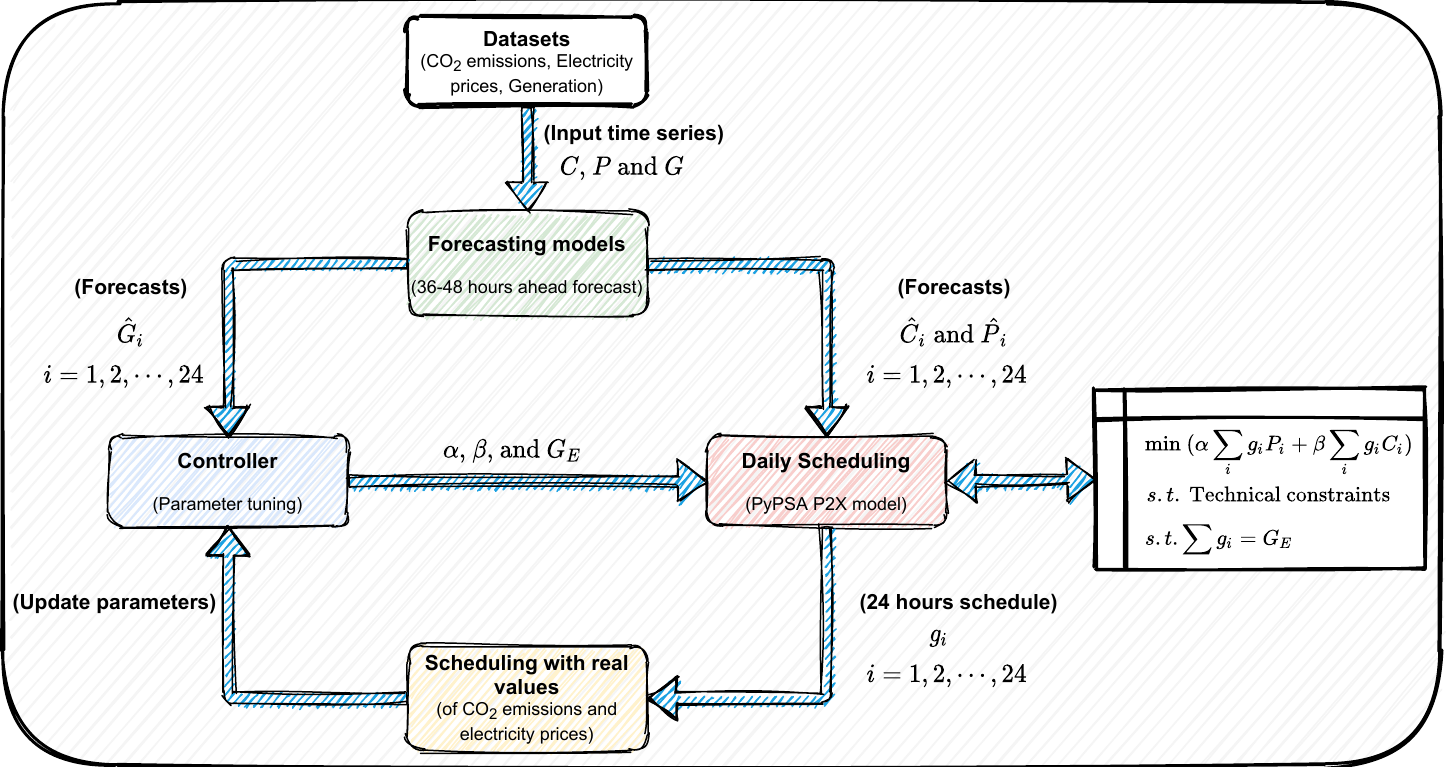}}
	\caption{Schematic of the proposed methodology.}
	\label{Fig_3}
\end{figure}

The dataset for the Danish scenario from ENSTO-E Transparency platform \cite{transparency} is used in this study. The spot prices with hourly resolutions are used directly from this platform, whereas the CO$_2$ intensity is calculated using the method of \cite{tranberg2019real}. This is a real-time carbon accounting method for European electricity markets, which applies the flow tracing concept to map power flow between an interconnected network of importing and exporting countries \cite{Tranberg2015}. This method uses local production mix, local power consumption, as well as imports and exports of electricity flow among neighboring countries that are also available in ENSTO-E Transparency platform. In this method, the hourly CO$_2$ intensity is found simultaneously for all areas by solving a set of linear equations. The imports from a neighboring area depend both on the mix within that area as well as the neighbors from which this area imports \cite{tranberg2019real}. The average CO$_2$ intensity per area per hour that is used in the following case studies, is derived from the specific CO$_2$ intensity per generation technology per area based on the ecoinvent 3.4 power plant database \cite{ecoinvent}. The result of using this flow tracing method is the historic CO$_2$ intensity in the grid. These historic values of CO$_2$ intensity can then be used to generate a forecast of the coming values. 

In the present study, both electricity prices and CO$_2$ intensity are forecasted with the ‘Method 1’ proposed in \cite{bokde281short}. This method is a short-term forecasting model based on the decomposition of the historic time series with moving average method and generation of three sub-series components named: trend, seasonal, and random ones. Figure 2 in \cite{bokde2020graphical} shows the block diagram of the moving average based decomposition used in the ‘Method 1’. The first step of this decomposition is the detection of the trend in the time series using the centered moving average method. Then the time series is detrended by eliminating the trend series from the original time series. The detrended series represents a seasonal component along with some noise values. Finally, the random component is extracted by subtracting trend and seasonal components from the original time series.

After decomposition, the trend and random components are forecasted with two different Autoregressive integrated moving average (ARIMA) models, whereas a Feed-forwards neural network (FFNN) model is used for the seasonal component. The addition of the forecasted values of these three components in the final forecasting results as shown in Figure 3 in \cite{bokde2020graphical}. 
Earlier two studies \cite{bokde281short, bokde2020graphical} confirm the suitability and higher accuracy of this method in short-term forecasting of electricity prices and CO$_2$ intensity time series. The forecast analysis and performance evaluation associated with the following case study is performed with the 'ForecastTB' tool \cite{forecasttb}.

\subsection{Short term scheduling}
The next step in the proposed methodology is the integration of power-to-X technologies with the energy market. Local production units have to supply the local demand while competing with other power units on the electricity market. In this study, three realistic technologies are modeled in the PyPSA tool \cite{PyPSA}. As previously mentioned the scheduling of units is performed by a scheduler managed by a controller. The scheduler is responsible for scheduling 24 hour periods, given the desired number of full load hours (FLH) to schedule the plant. The controller is used when a scheduling horizon longer than 24 hours is desired. Using historic data, the controller will, for every day, determine the number of FLH the scheduler should use. 

The short-term scheduler will receive the forecasted values of CO$_2$ intensities and electricity prices as input, along with the desired number of FLH of the plant. The number of FLH, thus, determines the production of the plant. The objective of the scheduler is to schedule the plant for the desired number of FLH while co-optimizing CO$_2$ intensities and the price of the electricity consumed. In Equation \eqref{e1}, the scheduler objective along with constraints is shown. The scheduler takes the form of a linear optimization problem and is implemented in the open-source tool \cite{PyPSA}. As technologies are considered as price takers in this paper, a single bid is determined for every hour in contrast to the more complex bidding curves sometimes used.

\begin{equation}
	\label{e1}
	\begin{aligned}
		\min_{g_i} \quad & \alpha \sum_{i=0}^{24}{g_{i,n}\hat{C}_{i,n}} + \beta \sum_{i=0}^{24}{g_{i,n}\hat{P}_{i,n}} \\
		\textrm{s.t.} \quad & \sum_{i=1}^{24}{g_{i,n}} = G_{E,n}\\
		\textrm{s.t.} \quad & g_{0,n} = g_{24,n-1}\\
		\textrm{s.t.} \quad & \beta = 1-\alpha\\
		\textrm{s.t.} \quad & \text{Technical constraints}\\
	\end{aligned}
\end{equation}
where, $\hat{C}_{i,n}$ and $\hat{P}_{i,n}$ are hourly forecasted values of CO$_2$ intensity and electricity price, $\alpha$ and $\beta$ are the weighting for $C_{i,n}$ and $P_{i,n}$ values, respectively. $g_{i,n}$ is the electricity purchase at $i^{th}$ hour on $n^{th}$ day, and $G_{E,n}$ is total electricity purchase for the $n^{th}$ day.

Using an optimizer, the amount of electricity to be purchased in every hour $g_{i,n}$ is found. By changing the $\alpha$ and $\beta$ weights, priority can be shifted between emission optimization and cost optimization. The above optimization is performed every day before bids are to be placed at 12:00 in the day-ahead market.

The cost function in Equation~\eqref{e1} is modified for the heat pump model (third technology) considering fixed gas prices and CO$_2$ emissions for the gas boiler as shown in \eqref{e1a}:

\begin{equation}
	\label{e1a}
	\begin{aligned}
		\min_{g_{i,n},b_{i,n}} \quad & \alpha \sum_{i=0}^{24}{(g_{i,n}\hat{C}_{i,n} + b_{i,n}C_{gas})} + \beta \sum_{i=0}^{24}{(g_{i,n}\hat{P}_{i,n} + b_{i,n}P_{gas})} \\
	\end{aligned}
\end{equation}
where, $C_{gas}$ and $P_{gas}$ are the fixed CO$_2$ emissions and gas prices for the gas boiler.

\subsection{Long term scheduling}
Scheduling can be performed with a daily, monthly, or yearly time horizon. The time horizon is the time the unit has to complete production for the desired number of FLH. In other words, for daily scheduling, the desired number of FLH for the given day, is specified every day by the user, and the scheduler is applied directly without any use of the controller. For monthly scheduling, the number of FLH over the course of a month is specified, and for yearly scheduling, the desired number of FLH for the entire year is specified, i.e. production on individual days can vary as long as they sum up to the required amount for the full month or year. 

For both monthly and yearly time horizons, the controller is implemented to determine the daily FLH to be scheduled. The controller will then consider historic CO$_2$ emission and electricity prices to determine how many hours to schedule for each day. For example, assuming that 6000 FLHs over the duration of a year are of user interest, the controller will distribute the FLHs over the year based on historic power prices and CO$_2$ intensity. The key challenge for the controller is to recognize on which days it is best to allocate production with out knowledge of the full month or year.


\subsubsection{Daily scheduling}
When using the daily scheduling time horizon the controller module is not implemented. Instead, the daily full load hours scheduling is performed with a straightforward procedure, where 38 hours ahead forecasted values are provided to the daily scheduler at 10.00 AM (when the market bidding process starts). This procedure of forecasting (for 38 hours ahead values of CO$_2$ intensity and electricity prices) and daily scheduling with scheduler continued for 365 days of the year, with the constraints shown in Equation~\eqref{e1}. The user must specify the FLHs $G_{E,n}$ along with $\alpha$ and $\beta$ for every day. 

\subsubsection{Monthly scheduling}
The monthly and yearly scheduling approaches are a bit different and computationally heavier than the daily ones. In the monthly scheduling, the number of hours to be scheduled with the daily scheduler is not fixed, as it was in the daily scheduling approach. Rather the user needs to employ some technique to estimate which are the ‘good’ or ‘bad’ days in the targeted month, in advance. In this paper, ‘good’ days are refereeing the market days that are responsible for carbon-efficient and cheaper electricity. For example, if a user wishes to schedule optimum hours each day in the month of January, then he/she must be aware of which are the ‘good’ days for scheduling/bidding the market and for how many hours he/she must bid in the ‘good’ day. This information is crucial for monthly scheduling since the users have the flexibility to choose more or less than the average ($\frac{FLH}{365}$) hours each day, which was not possible in the daily scheduling approach. However, the sum of the hours must be fixed for the whole month. 

To solve this problem, a controller is employed in the proposed methodology (as shown in Figure \ref{Fig_3}). The task of this controller is to suggest the number of hours to be scheduled by the daily scheduler. The controller is a separate entity that also uses a model similar to the ‘daily scheduler’, but for a longer scale, i.e., monthly or yearly period. 

For every day the controller has to determine the number of FLHs to schedule it will run through the following procedure every day at 10:00. 
The controller takes the linear optimization problem formulated in \eqref{e1} and expands it to span a full month instead of 24 hours. Combining CO$_2$ intensity and electricity price time series data for the last 28 days, the current day and the next day. More precisely historic data of 28 days and 10 hours and forecasted values for 38 hours is used. The desired number of FLH for the entire month is used as input for this optimization problem. The result is, thus, a schedule of the unit for the past 28 days plus the current and the next day. The planned number of FLH determined for the last day in the monthly model will then be used as input for the daily scheduler assuming that the past 30 days serves as a reasonable reference for the current month. The outcome of this procedure is a first estimate of the FLH to be scheduled the following day by the daily scheduler.




In monthly scheduling, it is expected that the average daily hours $\hat{G}_{E,n}$ over the span of a month should be $\frac{FLH}{\text{n days in month}}$. However, the procedure discussed so far does not ensure this in monthly and yearly scheduling since the number of hours scheduled each day is not fixed. Therefore, it can not be expected that the proposed generation $\hat{G}_{E,n}$ over the month should be equal to exactly the specified number of FLH. 


Ideally, the difference between the desired number of FLH and the scheduled FLH $\sum_{n=1}^{30}{\hat{G}_{E,n}}$ should be minimum. The controller in the proposed methodology attempts to minimize this difference with a balancing technique. This balancing is performed with function $f$ as shown in \eqref{e7}.

\begin{equation}
	\label{e7}
	f = \frac{\sum_{i=1}^{n}{\hat{G}_{E,i}} + \frac{(N-n)}{N}G_{E,M,perfect}}{G_{E,M,perfect}} 
\end{equation}
where, $\hat{G}_{E,i}$ is the estimated `good' hours of generation for $n^{th}$ day, and $G_{E,M,perfect}$ is the ideal number of generation hours that must be scheduled each month to achieve the FLH schedule for the year. For example, for $FLH=500$, $G_{E,M,perfect} = 500$. $N$ is the number of days in the particular month.

The ratio $f$ is expected to be almost 1 and used for updating generation for the next $(n+1)^{th}$ day with expression \eqref{e8}.

\begin{equation}
	\label{e8}
	\hat{G}_{E,n+1} \leftarrow \frac{\hat{G}_{E,n+1}}{f} 
\end{equation}

When $f < 1$, the next day generation will be increased by $\frac{1}{f}$ times, and when $f > 1$, it will be decreased by  $\frac{1}{f}$ times. For the long term scenario, this balancing technique attempts to maintain $f \approx 1$, and minimize the difference between $G_{E,M,perfect}$ and $\sum_{n=1}^{N}{\hat{G}_{E,n}}$. 
A sample graph indicating the behavior of $f$ throughout a year-long FLHs scheduling process is shown in Figure \ref{App1}.
Although, this technique does not ensure $FLH = \sum_{n=1}^{30}{\hat{G}_{E,n}}$, the difference between these parameters get minimized significantly.

Based on the balancing technique, the total generation for the targeted day ($\hat{G}_{E,n+1}$) is provided to the daily scheduler. The daily scheduler returns the optimum generation schedule for the targeted day (say, day 1 of the month). This process is repeated for the number of days in the targeted month and averaged over the period with expression \eqref{e4}.

\begin{equation}
	\label{e4}
	G_{E,m} = \frac{\sum_{n=1}^{\text{days in month `\textit{m}'}}{\hat{G}_{E,n}}}{\text{days in month `\textit{m}'}}
\end{equation}
where, $G_{E,m}$ is total generation for the $m^{th}$ month.


\subsubsection{Yearly scheduling}
A procedure similar to the monthly scheduling is used for the yearly FLH scheduling. The difference being the longer historic time series (one-year span, i.e. 365 days) of CO$_2$ intensity and electricity prices used in the optimization problem. 
As with monthly scheduling, the problem with the long time horizon is solved every day to determine the number of 'good' hours. This number of FLH is then given to the daily scheduler that will determine what hours to schedule the plant. 
Finally, the schedule returned by the daily scheduler is averaged over 365 days, which represents the averaged FLH generation based on the yearly scheduling ($G_{E,Y}$) with the expression \eqref{e9}.

\begin{equation}
	\label{e9}
	G_{E,Y} = \frac{\sum_{n=1}^{365}{\hat{G}_{E,n}}}{365}
\end{equation}
where, $G_{E,Y}$ is total generation for the whole year with the yearly scheduling approach.



\section{Results}
The proposed case study attempts to interface the local production units with the day-ahead electricity market and run them while competing with real-time electricity prices as well as maintaining carbon emissions levels as low as possible. For this purpose, the local production units are modeled considering the three different scenarios previously described. 
The single electrolyzer is a simple example of a generator that is interfaced in the first case of this study such that its marginal cost was a trade-off combination of electricity prices and CO$_2$ intensity as shown in expression (\ref{e1}). We have considered several combinations of weights assigned to these parameters by varying the values of $\alpha$ and $\beta$, such that $\alpha = 1- \beta \in \{0,0.1, \cdots ,0.9,1\}$, and scheduled the 6000 FLHs throughout the year.

\begin{figure}[p]
	\centering
	\begin{subfigure}{.49\textwidth}
		\centering
		\includegraphics[width=\textwidth]{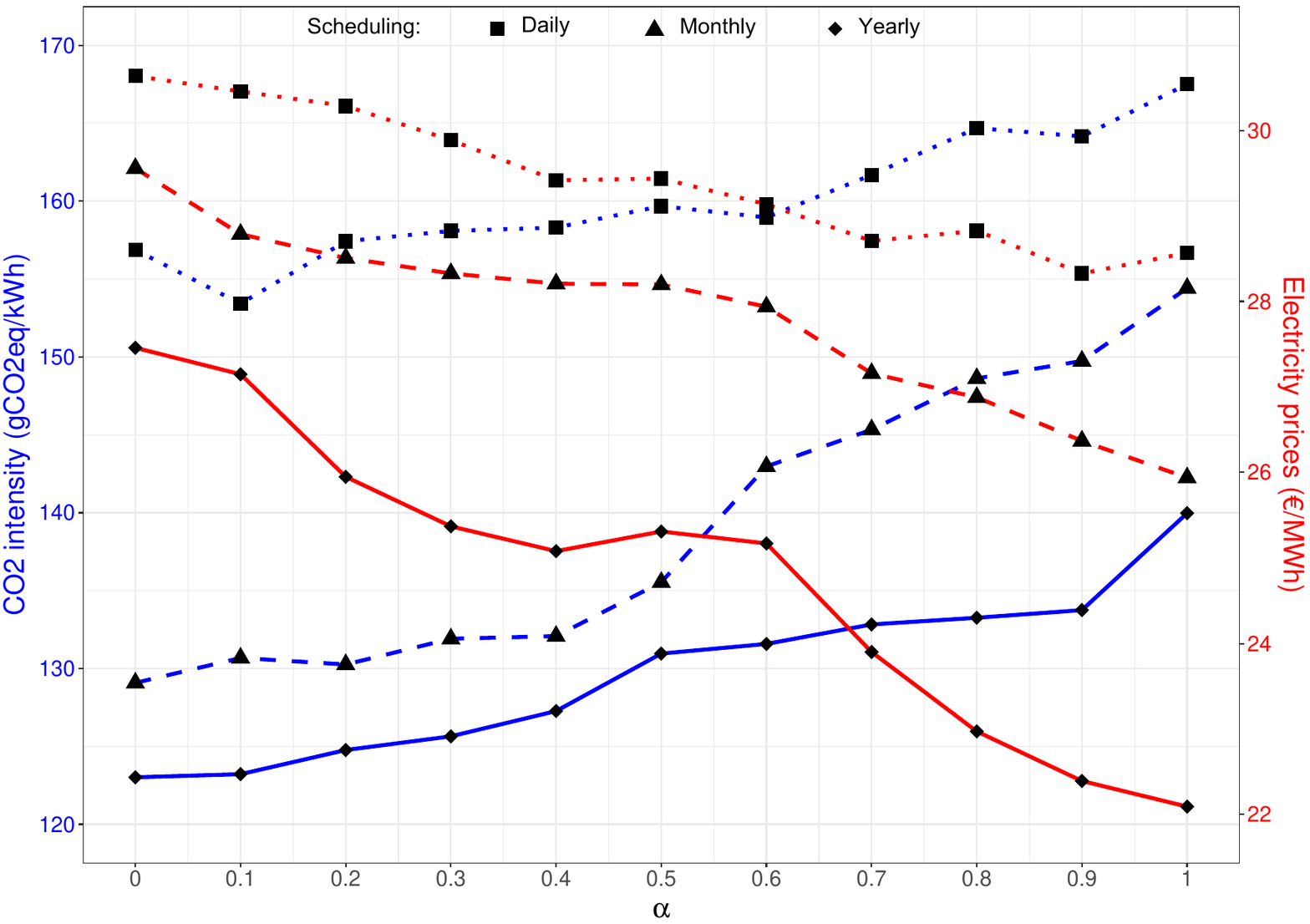}
		\caption{}
		\label{tradeoffa}
	\end{subfigure}
	\begin{subfigure}{.49\textwidth}
		\centering
		\includegraphics[width=\textwidth]{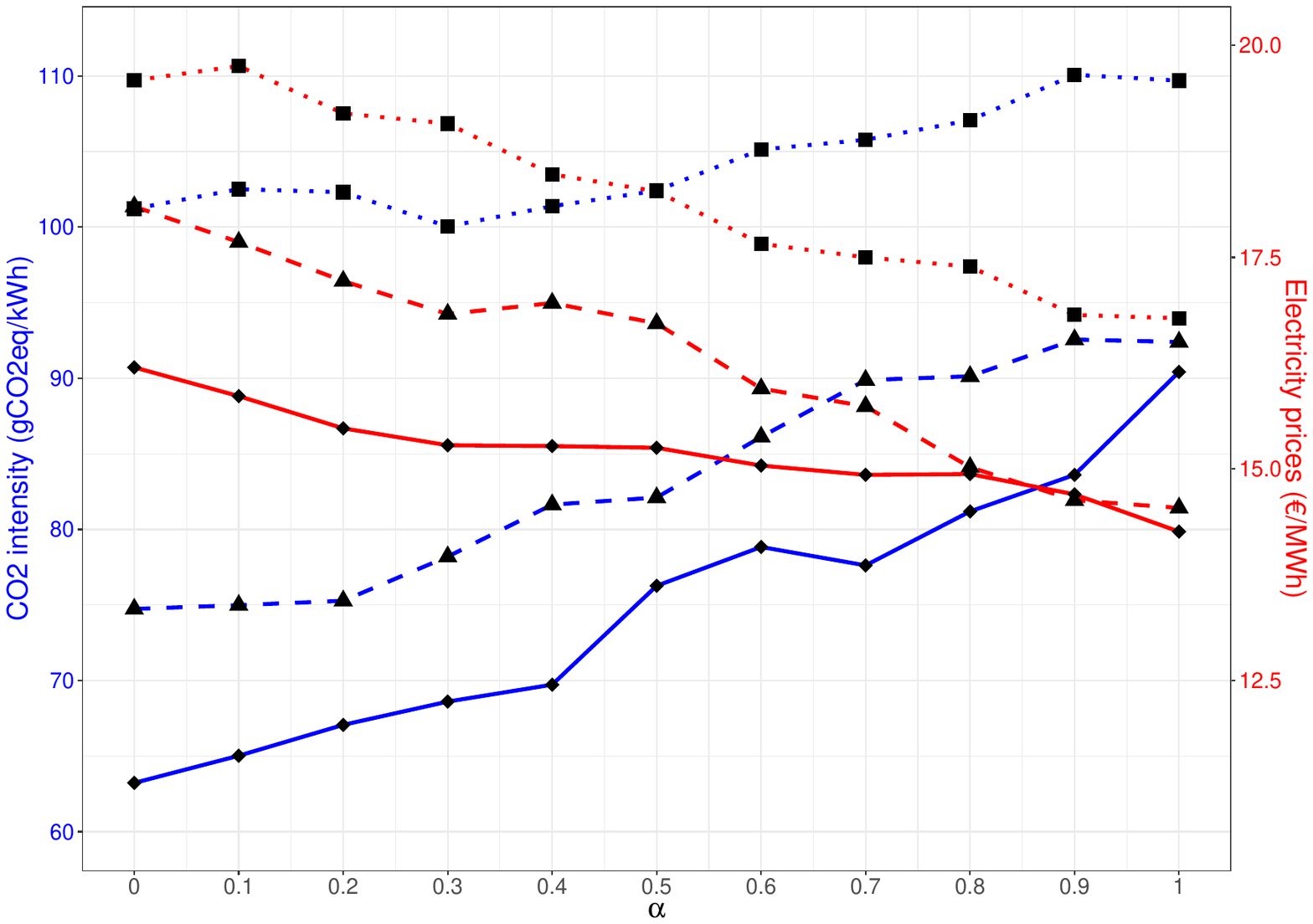}
		\caption{}
		\label{tradeoffb}
	\end{subfigure}
	\begin{subfigure}{.49\textwidth}
		\centering
		\includegraphics[width=\textwidth]{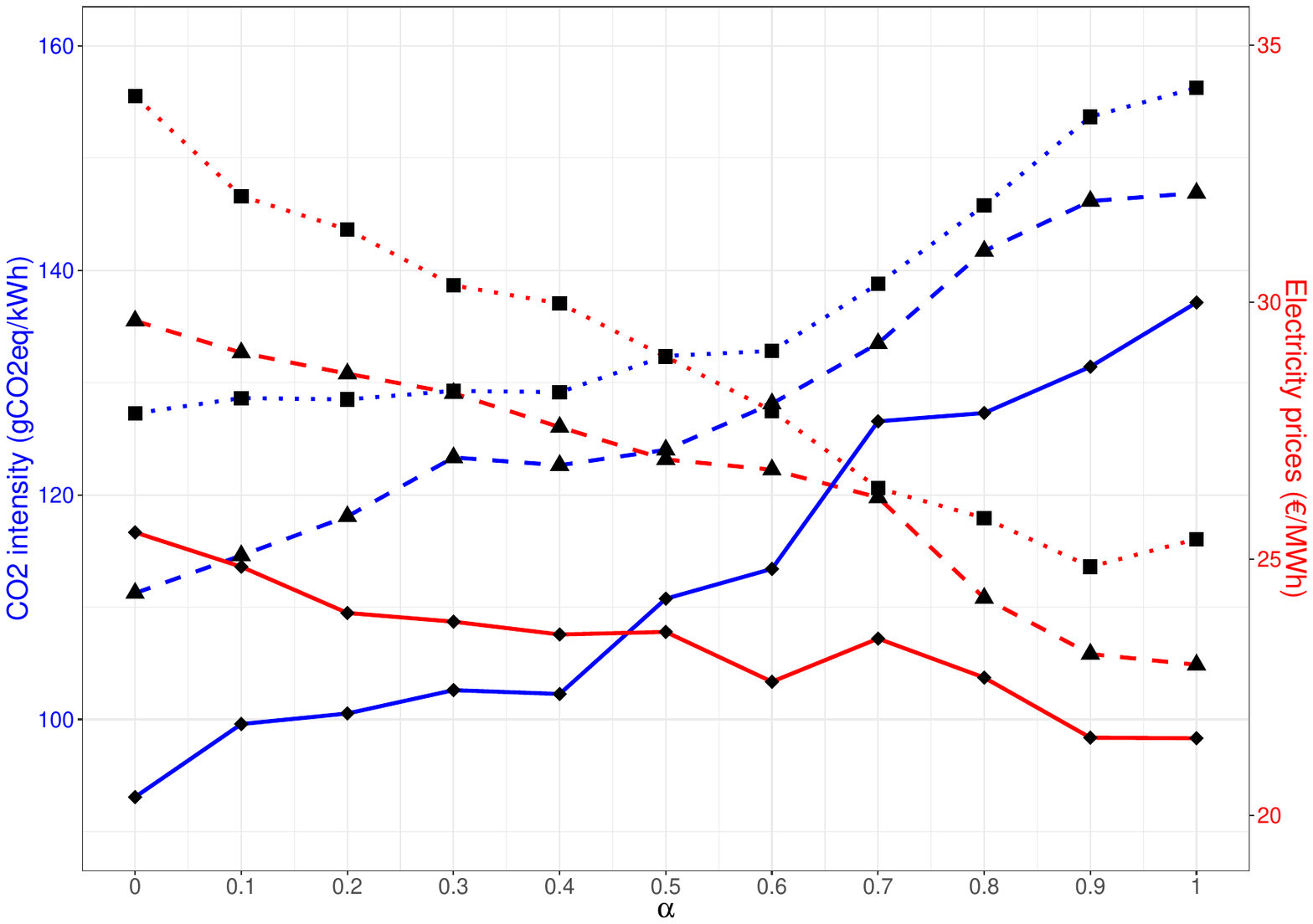}
		\caption{}
		\label{tradeoffc}
	\end{subfigure}
	\caption{Observed changes in the CO$_2$ intensity and electricity prices for 6000 FLHs with respect to change in wights for CO$_2$ intensity ($\alpha$) of the line for daily, monthly and yearly FLHs scheduling in technologies: a) Electrolyzer, b) Methanation plant, and c) Heat pump, respectively.}
	\label{tradeoff}
\end{figure}

Figure \ref{tradeoff} shows the electricity prices and CO$_2$ intensity for the generation with daily, monthly, and yearly FLH scheduling approaches for a) Electrolyzer, b) Methanation plant and c) Heat pump, respectively. 
These outcomes are well aligned with the scheduling results obtained with the graphical method in \cite{bokde2020graphical}. These results show the clear trade-off between electricity prices and CO$_2$ intensity when scheduled the generations with the proposed  model for electrolyzer unit, methanation, and heat pump plants. A clear nonlinear pattern is observed in the trade-off between cost and CO$_2$ intensity. 

The proposed methodology allows altering the ratio of CO$_2$ intensity and electricity prices by updating the $\alpha$ parameter. The effect of changing $\alpha$ on the scheduling of hours with the proposed methodology for the Western Denmark (DK-1) region for the years 2018-19. The hours are scheduled for daily, monthly, and yearly FLHs. In the shown results (in Figure \ref{tradeoff}), 16.43, 500, and 6000 FLHs are scheduled daily, monthly, and yearly, respectively. The FLHs in yearly scheduling are represented with a solid line, whereas that in monthly and daily scheduling are shown with dashed and dotted lines, respectively.

The sub-figures in Figure \ref{tradeoff} depict that the prices decreasing with increasing CO$_2$ emission intensity with the increasing values of alpha and vice-versa. This observation shows a trade-off between prices and CO$_2$ emission intensity since the lowest prices and intensity are not achievable at the same time. Besides, it can be observed that the magnitude of electricity prices and CO$_2$ emission intensity increases from yearly to monthly to daily scheduling.

These results are well-aligned with the trade-off obtained in the CO$_2$ emission intensity and electricity prices at the ideal forecast values as shown in Figure \ref{tradeoffideal}. 

\section{Conclusion}
Power-to-X has emerged as a promising source of flexibility in electricity systems with the increase in the share of variable renewable energy sources. The renewable energy sources are highly intermittent, therefore, practicing a proper scheduling mechanism is essential to make the power-to-X plants more economically viable. Apart from the economical aspects, the emissions (especially, CO$_2$ emission intensity) should be considered in the electricity market scheduling mechanism to achieve sustainable growth.

This study has introduced a power-to-X scheduling framework that is capable of co-optimizing CO$_2$ emission intensity and electricity prices in the day-ahead market scheduling. In the simulation study using historic electricity prices, CO$_2$ emission intensity, and their forecasts for the western Danish day-ahead market, three realistic models for flexible dispatch and its impact on the electricity market scheduling is observed. The proposed method has the potential to schedule any flexible demand plant, such as power-to-X units, heat pumps, or more complex systems composed of several technologies. We have investigated three different power-to-X technologies of varying complexities, i.e., stand-alone electrolyzer, a plant consisting of electrolyzer cascaded with a methanation reactor, and a heat pump combined with a gas boiler and thermal storage tank.

This study investigated the possible relations between CO$_2$ emission intensity and electricity prices while co-optimizing them, and observed a clear non-linear trade-off favoring a weighted optimization between the two objectives. It is observed that it is not possible to obtain the lowest price and CO$_2$ emission intensity at the same time, however, better scheduling strategy plans can be derived based on the observed trade-off analysis.

\appendices
\section{The parameters used for Power-to-X technologies.}
Tables \ref{T1}, \ref{T2}, \ref{T3} and \ref{T4} show the parameters used for Power-to-X technologies discussed in the paper.

\begin{table}[]
	\begin{minipage}{\linewidth}
		\tiny
		\centering
		\caption{The parameters used for electrolyzer plant in Technology 1.}
		\label{T1}
		\begin{tabular}{|c|c|c|c|}
			\hline
			\multicolumn{4}{|c|}{Generator:   Electrolyzer}                                                                                                                 \\ \hline
			Attribute                                                    & PyPSA variable                                                    & Used Value    & Unit         \\ \hline
			Efficiency                                                   & efficiency                                                        & 100           & \%           \\ \hline
			Part load                                                    & \begin{tabular}[c]{@{}c@{}}p\_min\_pu =\\ p\_max\_pu\end{tabular} & $g_{24,n-1}$\footnotemark[1]          & MW           \\ \hline
			Nominal power                                                & p\_norm                                                           & 1             & MW           \\ \hline
			Ramp rate up                                                 & ramp\_limit\_up                                                   & 0.3           & MW/hour      \\ \hline
			\begin{tabular}[c]{@{}c@{}}Ramp rate\\ down\end{tabular}     & ramp\_limit\_down                                                 & 0.3           & MW/hour      \\ \hline
			\begin{tabular}[c]{@{}c@{}}Ramp rate\\ start up\end{tabular} & ramp\_limit\_start\_up                                            & 0.15          & MW/hour      \\ \hline
			\begin{tabular}[c]{@{}c@{}}Minimum\\ up time\end{tabular}    & min\_up\_time                                                     & 2             & hours        \\ \hline
			Marginal cost                                                & marginal\_cost                                                    & weighted cost\footnotemark[2]  & currency/MWh \\ \hline
			Bus                                                          &                                                                   & H2            &              \\ \hline
		\end{tabular}\\
		\footnotemark[1]{Generation in last hour of the earlier day}\\
		\footnotemark[2]{Weighted cost shown in expression \eqref{e1}}
	\end{minipage}
\end{table}
\begin{table}[]
	\begin{minipage}{\linewidth}
		\tiny
		\centering
		\caption{PyPSA parameters used for Methanation plant in Technology 2.}
		\label{T2}
		\begin{tabular}{|c|c|c|c|}
			\hline
			\multicolumn{4}{|c|}{Generator: Electrolyzer}                                                                                                                        \\ \hline
			Attribute                                                           & PyPSA variable                                                    & Used Value    & Unit         \\ \hline
			Efficiency                                                          & efficiency                                                        & 70            & \%           \\ \hline
			Part load                                                           & \begin{tabular}[c]{@{}c@{}}p\_min\_pu =\\ p\_max\_pu\end{tabular} & $g_{24,n-1}$\footnotemark[1]          & MW           \\ \hline
			Nominal power                                                       & p\_norm                                                           & 6             & MW           \\ \hline
			Ramp rate up                                                        & ramp\_limit\_up                                                   & 0.3           & MW/hour      \\ \hline
			Ramp rate down                                                      & ramp\_limit\_down                                                 & 0.3           & MW/hour      \\ \hline
			Ramp rate start up                                                  & ramp\_limit\_start\_up                                            & 0.15          & MW/hour      \\ \hline
			Minimum up time                                                     & min\_up\_time                                                     & 2             & hours        \\ \hline
			Marginal cost                                                       & marginal\_cost                                                    & weighted cost\footnotemark[2] & currency/MWh \\ \hline
			\begin{tabular}[c]{@{}c@{}}Hydrogen storage\\ capacity\end{tabular} &                                                                   & 6             & MWh          \\ \hline
			Bus                                                                 &                                                                   & H2            &              \\ \hline
			\multicolumn{4}{|c|}{}                                                                                                                                                 \\ \hline
			\multicolumn{4}{|c|}{Generator: Methanation plant}                                                                                                                 \\ \hline
			Attribute                                                           & PyPSA variable                                                    & Used Value    & Unit         \\ \hline
			Efficiency                                                          & efficiency                                                        & 77            & \%           \\ \hline
			Ramp rate up                                                        & ramp\_limit\_up                                                   & 0.04\footnotemark[3]          & MW/hour      \\ \hline
			Ramp rate down                                                      & ramp\_limit\_down                                                 & 0\footnotemark[4]             & MW/hour      \\ \hline
			Ramp rate start up                                                  & ramp\_limit\_start\_up                                            & 0.01\footnotemark[5]          & MW/hour      \\ \hline
			Minimum up time                                                     & min\_up\_time                                                     & 2             & hours        \\ \hline
			Marginal cost                                                       & marginal\_cost                                                    & 0             & currency/MWh \\ \hline
			Bus                                                                 &                                                                   & CH4           &              \\ \hline
		\end{tabular}\\
		\footnotemark[1]{Generation in last hour of the earlier day}\\
		\footnotemark[2]{Weighted cost shown in expression \eqref{e1}}\\
		\footnotemark[3]{Plant ramp-up time, hot start = 1 hour} \\
		\footnotemark[4]{Plant shut-down time = less than a minute} \\
		\footnotemark[5]{Plant ramp-up time, cold start = 0.24 hour} 
	\end{minipage}
\end{table}
\begin{table}[]
	\begin{minipage}{\linewidth}
		\tiny
		\centering
		\caption{PyPSA parameters used for Heat Pump in Technology 3.}
		\label{T3}
		\begin{tabular}{|c|c|c|c|}
			\hline
			\multicolumn{4}{|c|}{Generator: Heat Pump}                                                                                                                                                               \\ \hline
			Attribute                                                    & PyPSA variable                                                    & Used Value                                               & Unit         \\ \hline
			\begin{tabular}[c]{@{}c@{}}Efficiency\\ (COP)\end{tabular}   & efficiency                                                        & 300 (COP=3.0)                                                      &  \%            \\ \hline
			Part load                                                    & \begin{tabular}[c]{@{}c@{}}p\_min\_pu =\\ p\_max\_pu\end{tabular} & $g_{24,n-1}$\footnotemark[1]                                                  & MW           \\ \hline
			Nominal power                                                & p\_norm                                                           & 1                                                        & MW           \\ \hline
			Ramp rate up\footnotemark[2]                                                & ramp\_limit\_up                                                   & 0                                                        & MW/hour      \\ \hline
			\begin{tabular}[c]{@{}c@{}}Ramp rate\\ down\footnotemark[3]\end{tabular}     & ramp\_limit\_down                                                 & 0.25                                                     & MW/hour      \\ \hline
			\begin{tabular}[c]{@{}c@{}}Ramp rate\\ start up\end{tabular} & ramp\_limit\_start\_up                                            & 0                                                        & MW/hour      \\ \hline
			\begin{tabular}[c]{@{}c@{}}Minimum\\ up time\end{tabular}    & min\_up\_time                                                     & 2                                                        & hours        \\ \hline
			Marginal cost                                                & marginal\_cost                                                    & \begin{tabular}[c]{@{}c@{}}weighted\\ costs\footnotemark[4]\end{tabular} & currency/MWh \\ \hline
			Bus                                                          &                                                                   & Heat                                                     &              \\ \hline
		\end{tabular}\\
		\footnotemark[1]{Generation in last hour of the earlier day}\\
		\footnotemark[2]{Warm start-up time (hours) = 0 }\\
		\footnotemark[3]{Cold start-up time (hours) = 6}\\
		\footnotemark[4]{Weighted cost shown in expression \eqref{e1}}
	\end{minipage}
\end{table}
\begin{table}[]
	\begin{minipage}{\linewidth}
		\tiny
		\centering
		\caption{PyPSA parameters used for gas boiler in Technology 3.}
		\label{T4}
		\begin{tabular}{|c|c|c|c|}
			\hline
			\multicolumn{4}{|c|}{Generator: Boiler}                                                                                                                                  \\ \hline
			Attribute                                                    & PyPSA variable                                                    & Used Value             & Unit         \\ \hline
			Efficiency                                                   & efficiency                                                        & 90                     & \%           \\ \hline
			Part load                                                    & \begin{tabular}[c]{@{}c@{}}p\_min\_pu =\\ p\_max\_pu\end{tabular} & $g_{24,n-1}$\footnotemark[1]                & MW           \\ \hline
			Nominal power                                                & p\_norm                                                           & 1                      & MW           \\ \hline
			Ramp rate up\footnotemark[2]                                                   & ramp\_limit\_up                                                   & 0.004                  & MW/hour      \\ \hline
			\begin{tabular}[c]{@{}c@{}}Ramp rate\\ down\footnotemark[3]  \end{tabular}     & ramp\_limit\_down                                                 & 0.016                  & MW/hour      \\ \hline
			\begin{tabular}[c]{@{}c@{}}Ramp rate\\ start up\end{tabular} & ramp\_limit\_start\_up                                            & 0.1                    & MW/hour      \\ \hline
			\begin{tabular}[c]{@{}c@{}}Minimum\\ up time\end{tabular}    & min\_up\_time                                                     & 2                      & hours        \\ \hline
			Marginal cost                                                & marginal\_cost                                                    & $(\alpha \times C_f)+(\beta \times P_f)$\footnotemark[4]   & currency/MWh \\ \hline
			Bus                                                          &                                                                   & Heat                   &              \\ \hline
		\end{tabular}\\
		\footnotemark[1]{Generation in last hour of the earlier day}\\
		\footnotemark[2]{Warm start-up time (hours) = 0.1 }\\
		\footnotemark[3]{Cold start-up time (hours) = 0.4}\\
		\footnotemark[4]{$C_f = 201$ gCO2/kWh and $P_f = 20.1$ \euro{}/MWh}
	\end{minipage}
\end{table}

\section{Behavior of ratio `$f$'}
\begin{figure}[]
	\centering{\includegraphics[width=0.5\textwidth]{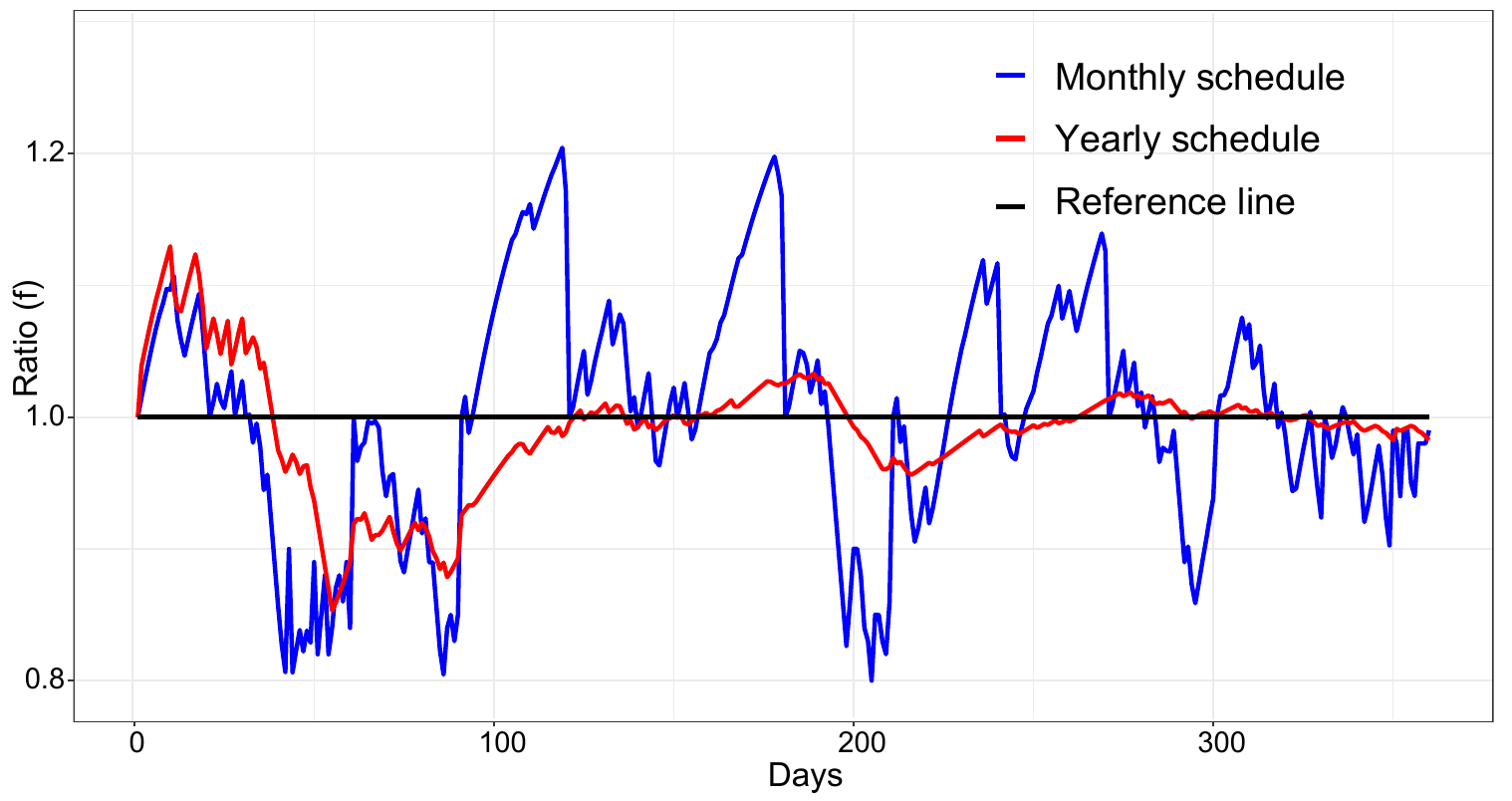}}
	\caption{Pattern of ratio $(f)$ throughout a yearly and monthly FLHs scheduling process.}
	\label{App1}
\end{figure}

Figure \ref{App1} indicates the behavior of ratio $f$ throughout a year-long FLHs scheduling process. This plot is representing the ratio $f$ for 6000 FLHs yearly scheduling for Electrolyzer plant, considering $\alpha = \beta = 0.5$.

\section{Illustration of the construction of models}
\begin{figure}[]
	\centering
	\begin{subfigure}{.35\textwidth}
		\centering
		\includegraphics[width=\textwidth]{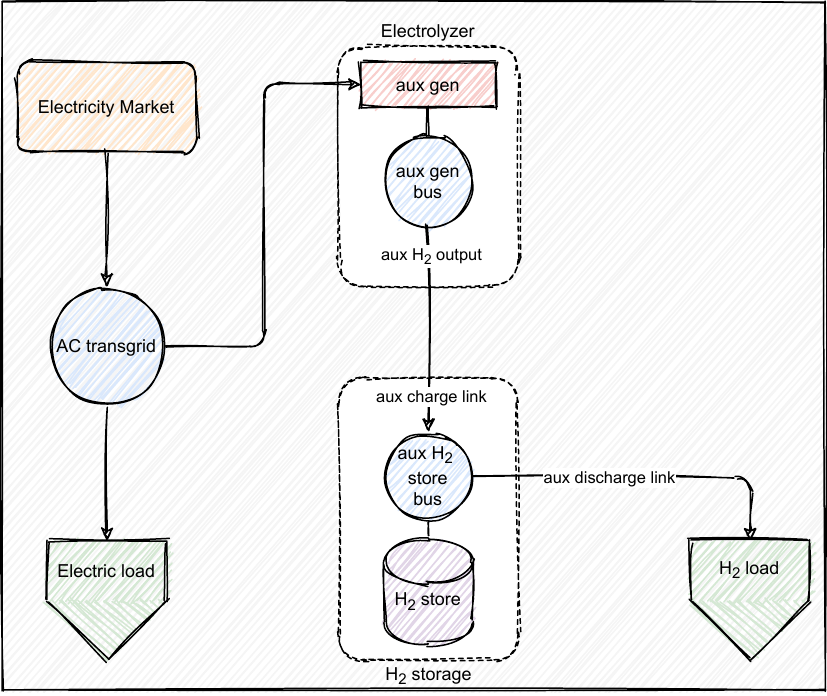}
		\caption{}
		\label{figxa}
	\end{subfigure}
	\begin{subfigure}{.35\textwidth}
		\centering
		\includegraphics[width=\textwidth]{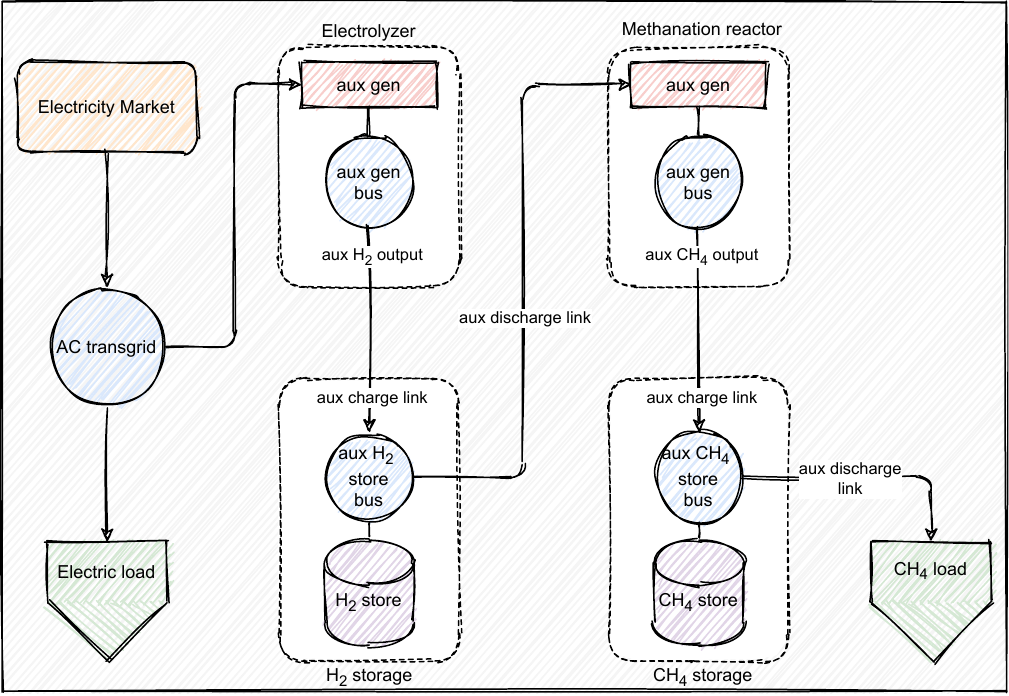}
		\caption{}
		\label{figxb}
	\end{subfigure}
	\begin{subfigure}{.35\textwidth}
		\centering
		\includegraphics[width=\textwidth]{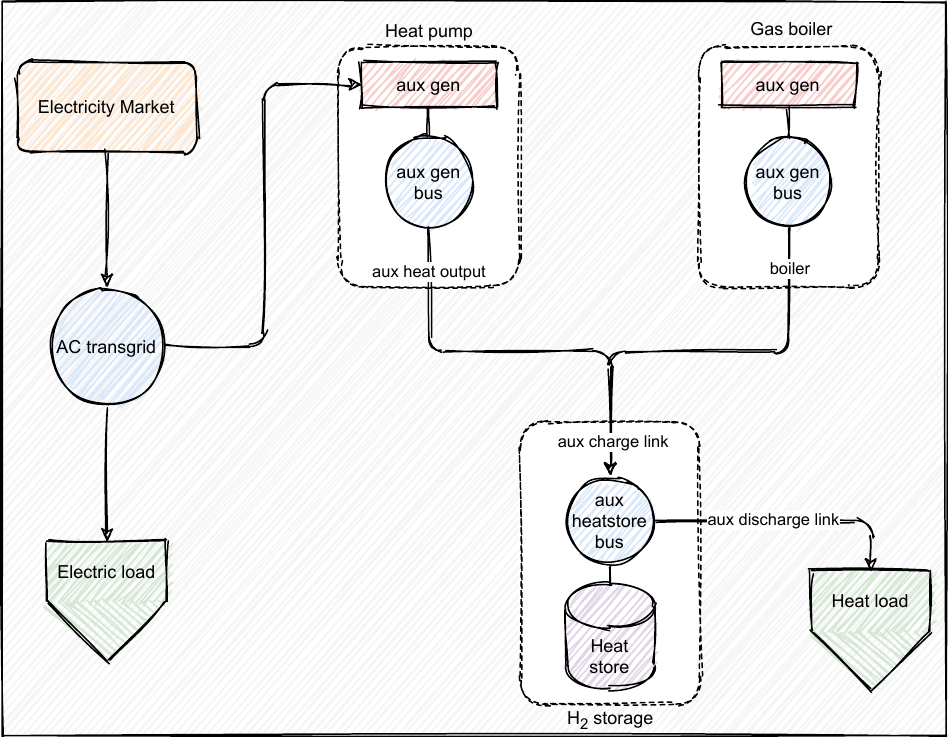}
		\caption{}
		\label{figxc}
	\end{subfigure}
	\caption{Illustration of the construction of the reference model network for a) Electrolyzer, b) Methanation plant, and c) Heat pump, respectively.}
	\label{illustration}
\end{figure}

Figure \ref{illustration} shows the Illustration of the construction of the reference model network for a) Electrolyzer, b) Methanation plant, and c) Heat pump, respectively. Each production unit is given within the dashed boxes and named with bold font. Squares indicate PyPSA generators, circles indicate buses and arrows with text indicates links.

\section{Trade-off with ideal forecasts}
\begin{figure}[]
	\centering
	\begin{subfigure}{.4\textwidth}
		\centering
		\includegraphics[width=\textwidth]{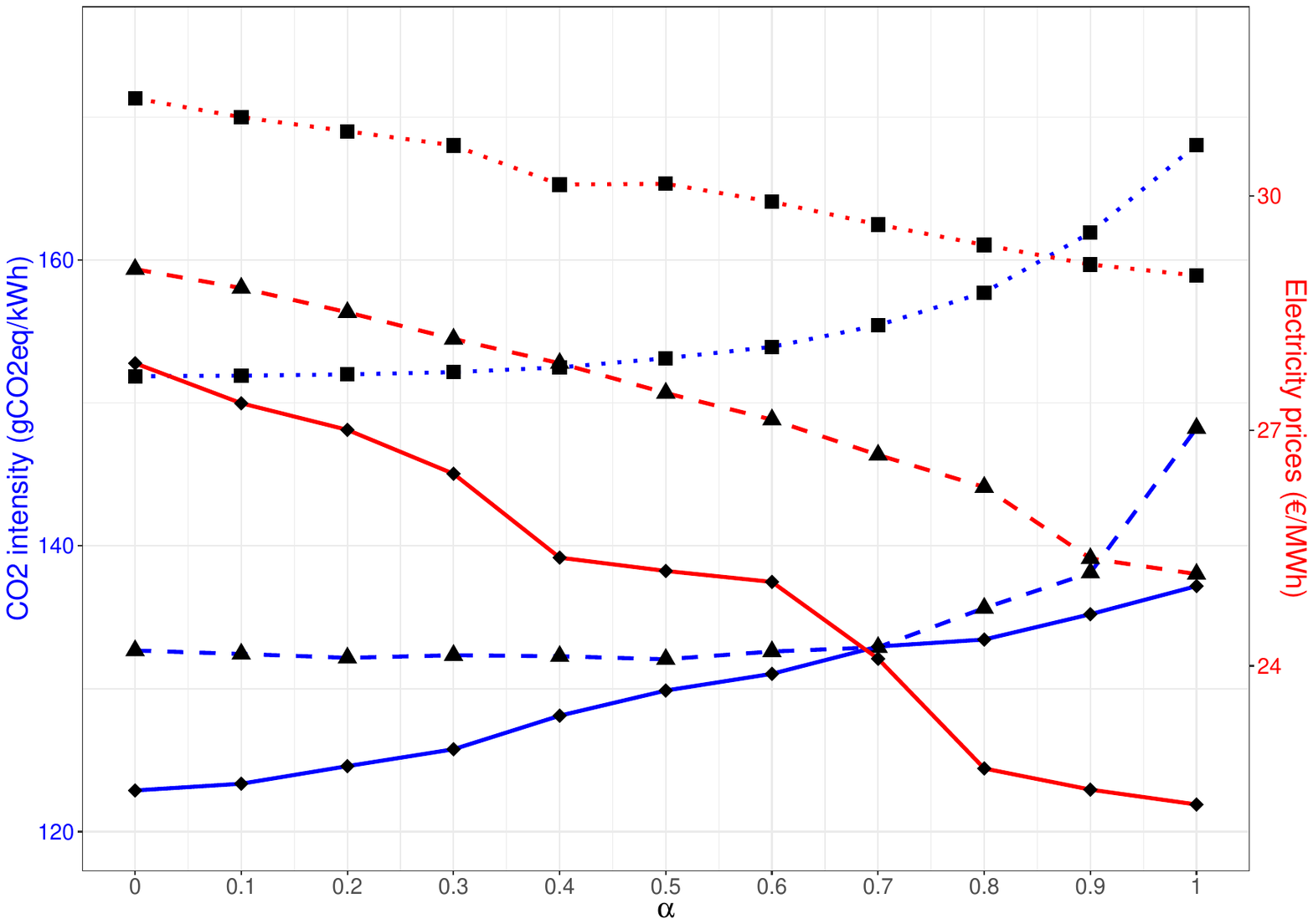}
		\caption{}
		\label{tradeoffaideal}
	\end{subfigure}
	\begin{subfigure}{.4\textwidth}
		\centering
		\includegraphics[width=\textwidth]{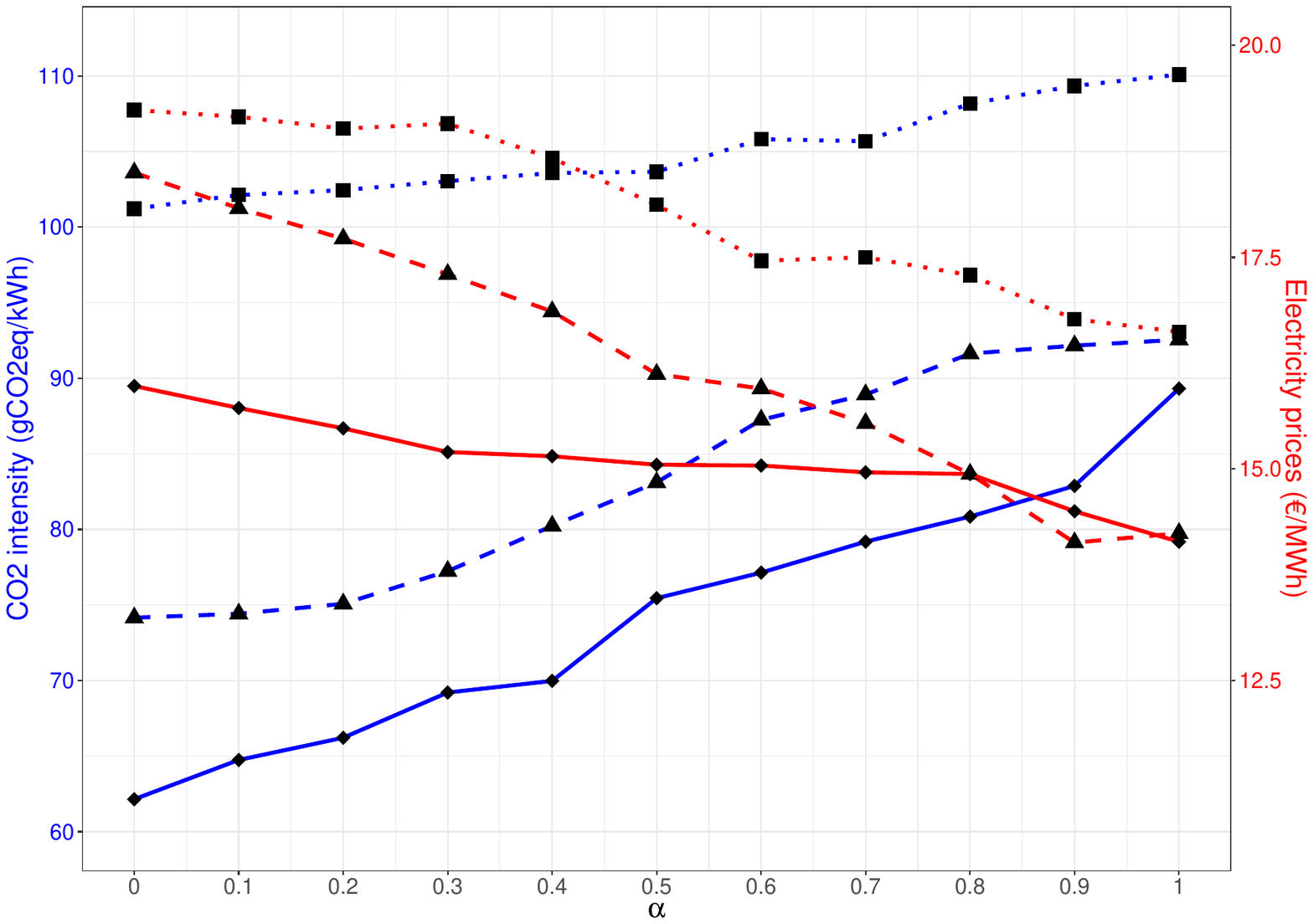}
		\caption{}
		\label{tradeoffbideal}
	\end{subfigure}
	\begin{subfigure}{.4\textwidth}
		\centering
		\includegraphics[width=\textwidth]{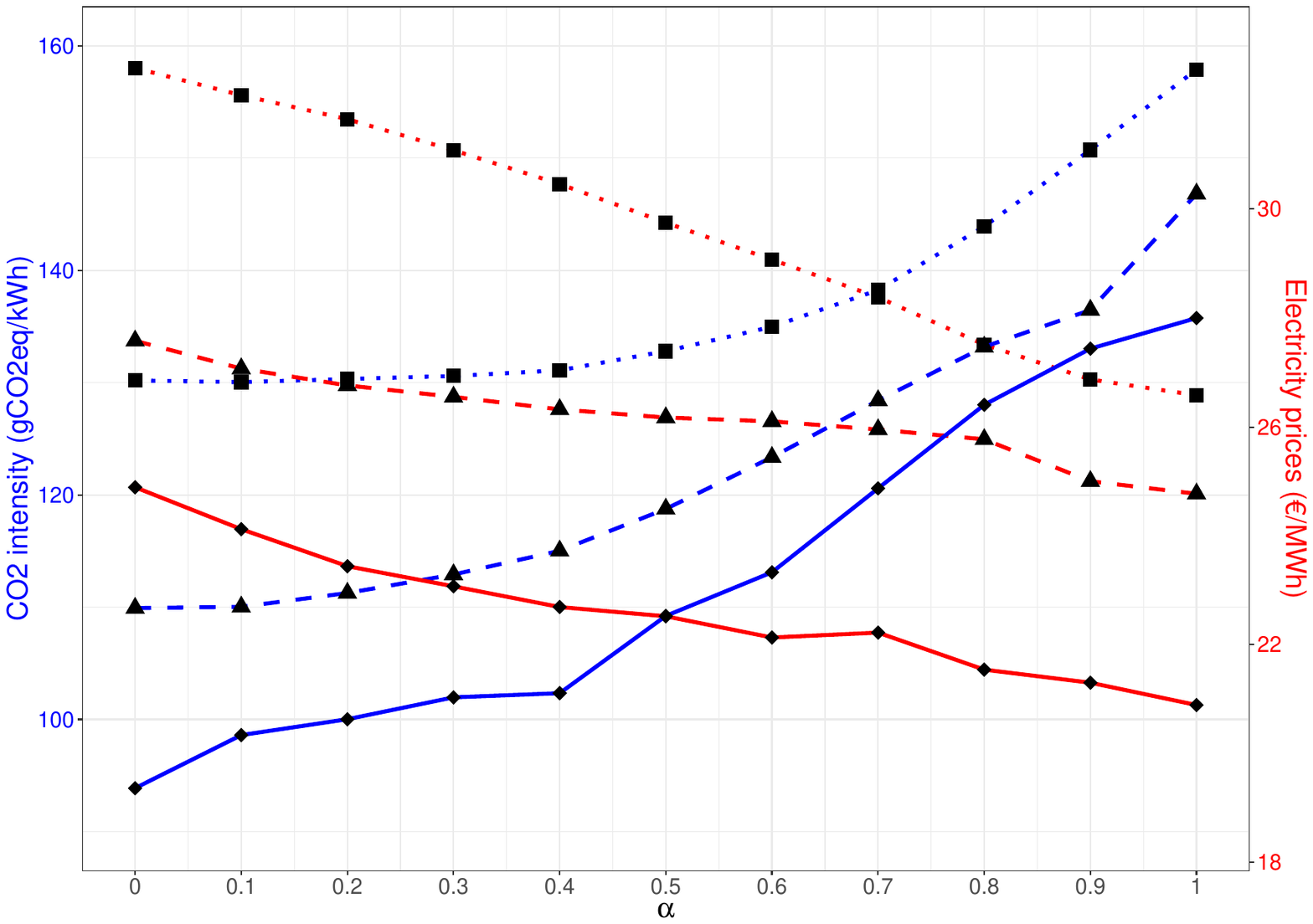}
		\caption{}
		\label{tradeoffcideal}
	\end{subfigure}
	\caption{Observed changes in the CO$_2$ intensity and electricity prices for 6000 FLHs with respect to change in wights for CO$_2$ intensity ($\alpha$) of daily, monthly and yearly FLHs scheduling with the ideal forecast values in technologies: a) Electrolyzer, b) Methanation plant, and c) Heat pump, respectively.}
	\label{tradeoffideal}
\end{figure}

Figure \ref{tradeoffideal} shows the observed changes in the CO$_2$ intensity and electricity prices for 6000 FLHs with respect to change in wights for CO$_2$ intensity ($\alpha$) of daily, monthly and yearly FLHs scheduling with the ideal forecast values in technologies: a) Electrolyzer, b) Methanation plant, and c) Heat pump, respectively. 

\section*{Acknowledgment}
This study was funded by Apple Inc. as part of the APPLAUSE bioenergy
collaboration with Aarhus University, Denmark. 

\clearpage



\bibliographystyle{IEEEtran}
\bibliography{mybibfile}
%
%
%

%
 \vskip -2\baselineskip plus -1fil 
\begin{IEEEbiographynophoto}{Neeraj Dhanraj Bokde}
is working as a Postdoctoral Researcher at Aarhus University, Aarhus, Denmark. He received a Ph.D. in data science from the Visvesvaraya National Institute of Technology, Nagpur, India. His major research contributions are in the domain of data science topics, focused majorly on time series analysis, software package development, and prediction applications in renewable energy. He is serving in editorial positions in Data in Brief, Frontiers in Energy Research, Energies, and Information journals.
\end{IEEEbiographynophoto}
 \vskip -2\baselineskip plus -1fil 
\begin{IEEEbiographynophoto}{Tim T. Pedersen}
is a PhD student at Aarhus University, Department of mechanical and production engineering  (MPE), in the Renewable Energy and Thermodynamics group. His research focuses on improving the energy system optimization models used for analysis  in the green transition of our energy supply. Using modern Modeling to Generate Alternatives (MGA) algorithms, Tim, studies near-optimal model solutions to uncover flexibility in the green transition of our energy supply. His current research, studies the technical limitations of ensuring a just energy transition of the European energy supply. Tim  has expertise within the fields of convex optimization, uncertainty analysis, energy system modeling, and data analysis.
\end{IEEEbiographynophoto}
 \vskip -2\baselineskip plus -1fil 
\begin{IEEEbiographynophoto}{Gorm B. Andresen}
PhD, is associate professor at AU-ENG where he leads the Renewable Energy and Thermodynamics research group (Twitter: @NRGYdynamics). The research of the group is concerned with the use of storage, transmission networks and energy conversion technology to facilitate integration of wind and solar energy in the energy system at all scales from continental to national and cities. This includes the effect of climate change on renewable energy systems. He lectures in theoretical and experimental courses in the domain of thermodynamics, flow and turbomachinery and renewable energy technologies and systems. The research targets high-impact international journals and is often executed in collaboration with industrial and public partners as well as internationally leading academic institutions. He holds a Ph.d. in experimental antihydrogen physics and co-authored the paper Trapped Antihydrogen published in Nature (2010).
\end{IEEEbiographynophoto}




\end{document}